\documentclass[prl,aps]{revtex4}

\usepackage{graphicx}
\usepackage{dcolumn}
\usepackage{bm}
\usepackage{epstopdf}

\begin{document}


\title{Low Temperature Susceptibility of the Noncentrosymmetric Superconductor CePt$_3$Si}

\author{D.P. Young, M. Moldovan, X.S. Wu, and P.W. Adams}
\affiliation{Department of Physics and Astronomy\\Louisiana State University\\Baton Rouge, Louisiana,
70803}%

\author{Julia Y. Chan}
\affiliation{Department of Chemistry\\Louisiana State University\\Baton Rouge, LA, 70803}%

\date{\today}

\begin{abstract}
We report ac susceptibility measurements of polycrystalline CePt$_3$Si down to 60 mK and in applied fields up to 9 T.  In zero field, a full Meissner state emerges at temperatures $T/T_c<0.3$, where $T_c=0.65$ K is the onset transition temperature.  Though transport measurements show a relatively high upper critical field $B_{c2}\sim 4-5$ T, the low temperature susceptibility, $\chi'$, is quite fragile to applied field, with $\chi'$ diminishing rapidly in fields of a few kG.  Interestingly, the field dependence of $\chi'$ is well described by the power law, $4\pi\chi'=(B/B_c)^{1/2}$, where $B_c$ is the field at which the onset of resistance is observed in transport measurements.  
\end{abstract}

\pacs{74.70.Tx,74.25.Ha,74.62.Bf}
\maketitle

	Two separate, but not unrelated considerations are paramount in the search for superconductors with extraordinary electronic properties.  The first is the role of crystal structure in determining the order parameter symmetry and the second is the nature of underlying electron correlations.  As an example of the latter, the newly discovered itinerant ferromagnetic superconductors, represented by  UGe$_2$ \cite{UGe2} and ZrZn$_2$ \cite{ZrZn2}, are believed to exhibit a superconducting phase that is, in fact, mediated by
a preemptive magnetic ordering.  Antiferrromagnetic analogs are also known, and include the quasi-2D heavy fermion 1-1-5 family, CeMIn$_5$, where M=Ir, Rh, or Co \cite{CeCoIn5}.  The superconducting phases in these two classes of systems are believed to be unconventional and are currently  the subject of intense investigation \cite{CeRhIn5}.    More recently a new heavy fermion superconductor was reported in the itinerant antiferromagnet CePt$_3$Si \cite{CePt3Si}, which owing to its magnetic ground state, falls broadly into the same class as the 1-1-5 superconductors.  The superconducting state in this material, however, is not only of interest in terms of magnetism, but also because it has a rather unusual crystal structure that lacks a center of inversion.    A considerable amount of theoretical work has been done on  two-dimensional (2D) systems lacking inversion symmetry, where the presence of a surface almost trivially guarantees the breaking of the symmetry \cite{Inversion}.  In such systems spin-orbit scattering can have a significant impact on the spin states of the superconductor, forcing spins to lie in the plane, thereby destroying the spin isotropy.  More generally, the effects of spin orbit scattering are in some sense enhanced in noncentrosymmetric systems, and will mix order parameter states of different parity.  Similar arguments have been for made bulk systems with non-inversion symmetry \cite{Inversion2}.  This suggests that CePt$_3$Si is a compelling candidate for the realization of nonconventional superconductivity mediated by the interplay between broken inversion symmetry, spin-orbit scattering, and magnetism  \cite{MnSi}.  Up to now only heat capacity and transport studies have been reported on CePt$_3$Si, though the antiferromagnetic order has been established via neutron scattering \cite{Neutron}.  In the present Letter, we present low temperature magnetic susceptibility measurements of polycrystalline CePt$_3$Si samples and show that the material exhibits a full Meissner state well below the superconducting transition.  Though believed to be a type II superconductor,  the field dependence of the zero temperature susceptibility obeys a power-law field dependence which is inconsistent with standard type II behavior but is perhaps more characteristic of quantum criticality \cite{QPT}.  Furthermore, we show that upon doping La for Ce at the $2\%$ level the superconducting transition is suppressed to below 50 mK.   The implications of these observations for the nature of the pairing is discussed.  

	Polycrystalline samples were prepared by argon arc melting of stoichiometric quantities of high purity Ce, Pt, and Si.  The resulting buttons were then annealed under vacuum at 900 $^o$C for $5-10$ days.  The correct stoichiometry was verified via standard powder x-ray diffraction analysis.  Transport measurements were made using a lock-in amplifier in a 4-wire configuration.  Low temperature ac susceptibility measurements were made with an astatically wound susceptibility coil operating at 505 Hz.  Typical excitation fields were $\sim0.05$ G, and external fields up to 9 T where applied parallel to the probe field.   Transport and susceptibility measurements below 2 K were carried out on a dilution refrigerator with a base temperature of 50 mK.  Measurements above 2 K were made using a Quantum Design PPMS magnetometer. 

\begin{figure}
\includegraphics[width=6in]{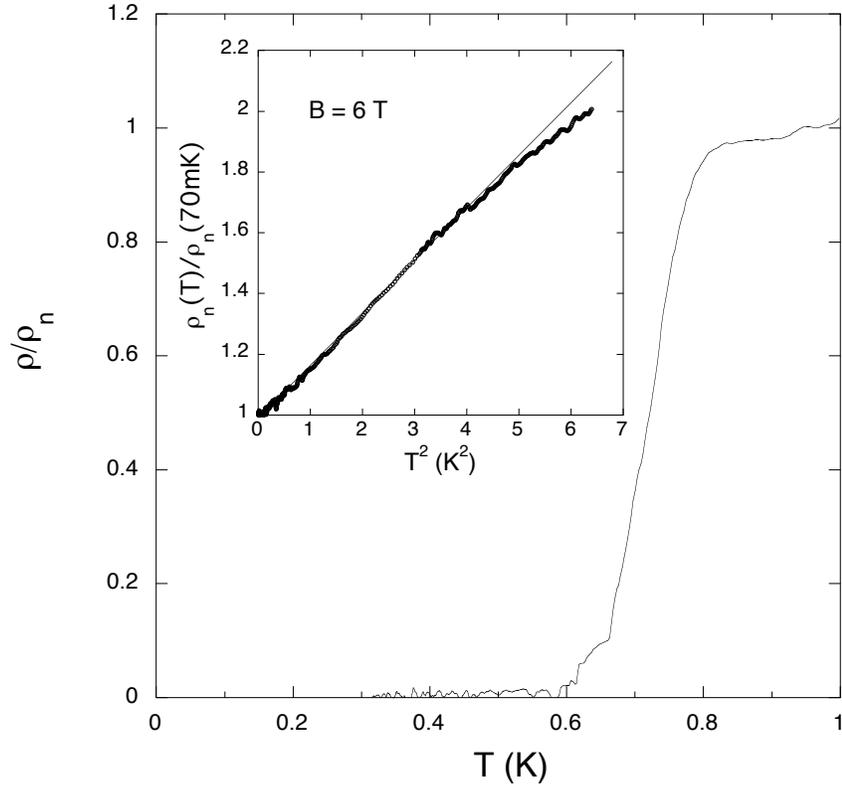}
\caption{\label{Rho-T} Relative resistivity of CePt$_3$Si as a function of temperature.  The midpoint of the superconducting transition is at $T_c=0.72$ K.  Inset:  quadratic temperature dependence of the normal state resistivity. }
\newpage
\end{figure}

	In Fig.\  \ref{Rho-T} we plot the normalized resistivity of CePt$_3$Si as a function of temperature in zero magnetic field.   The midpoint transition temperature $T_c=0.72$ K and transition width $\Delta T_c/T_c\sim0.15$ are in good agreement with the values reported in Ref.\  \cite{CePt3Si} .  In the inset of Fig.\ \ref{Rho-T} we plot the low temperature normal state resistivity as a function of $T^2$.   This data was taken in an applied field of 6 T to suppress the superconducting state and the linearity of the data is consistent with a Fermi liquid state.  The solid line has slope 0.167 K$^{-2}$ which is about a factor of three smaller than reported in Ref.\  \cite{CePt3Si}.  However the data in Ref.\ \cite{CePt3Si} were restricted to temperatures above $T_c$.    
	
\begin{figure}
\includegraphics[width=6in]{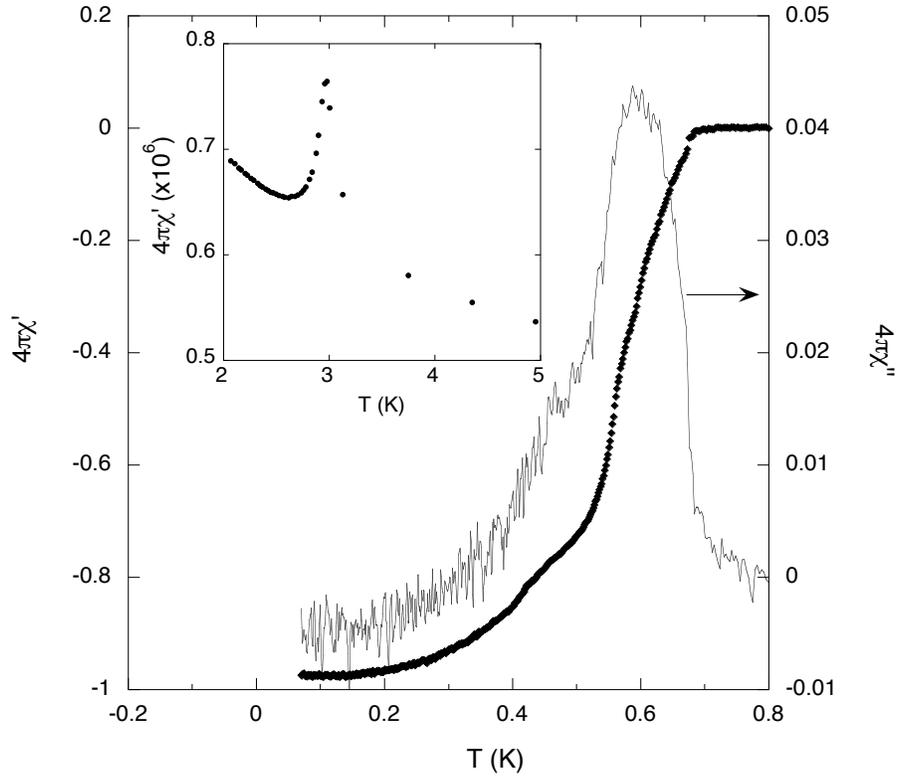}
\caption{\label{Chi-T}  Real and imaginary components of the ac susceptibility as a function of temperature in zero field.  Note that a full Meissner state is formed below $\sim0.3$ K.  Inset:  ac susceptibility as a function of temperature showing an antiferromagnetic transition at $T_N=2.97$ K.}
\newpage
\end{figure}

	The real and imaginary components of the ac susceptibility, $\chi'$ and $\chi''$ respectively, are plotted as a function of temperature in the main panel of Fig.\ \ref{Chi-T}.  The diamagnetic response of $\chi'$ and corresponding dissipation peak in $\chi''$ represent the superconducting transition.  A bulk superconducting state is observed, as $4\pi\chi'\rightarrow-1$ upon zero field cooling. The small local maximum in $\chi'$ between 0.4 K and 0.5 K was evident in all of the samples studied, but  its origin is unknown.   In the inset of Fig.\ \ref{Chi-T} we present $\chi'$ data above 2 K where there is clearly a magnetic ordering transition at $T_N=2.97$ K, which we believe is the N\'eel transition.  The $T_N$ in Fig.\ \ref{Chi-T} is moderately higher than that obtained from the heat capacity data of Ref.\ \cite{CePt3Si}.
	
\begin{figure}
\includegraphics[width=6in]{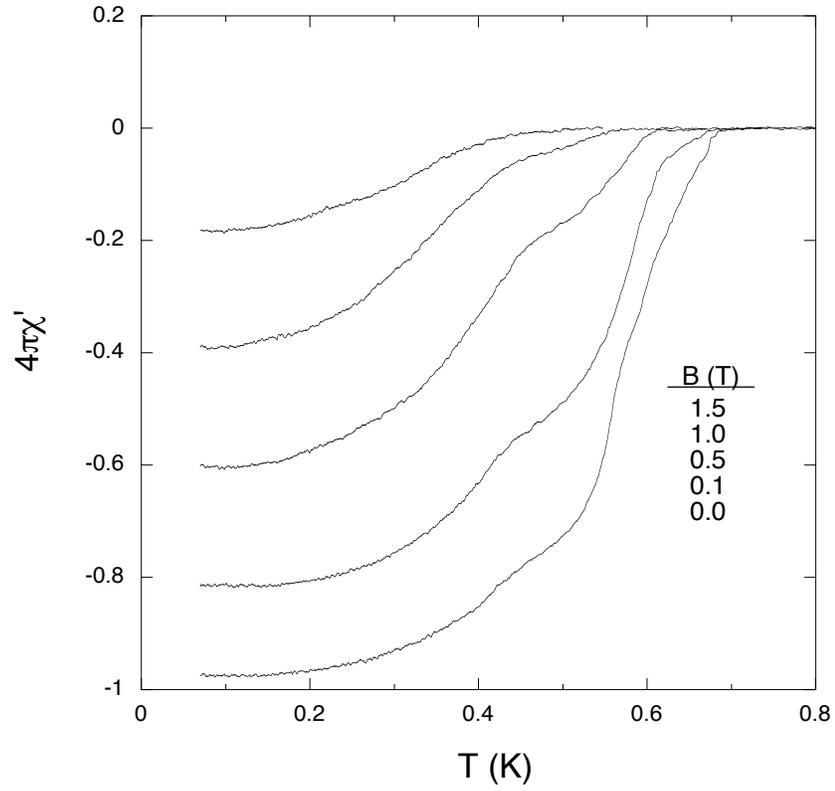}
\caption{\label{Chi-T-B} Susceptibility as a function of temperature in an external magnetic field.  Note the rapid increase in $4\pi\chi'$ upon the application of relatively modest fields.  The field at which the diamagnetic response was completely attenuated was $B_c\sim2.4$ T.}
\newpage
\end{figure}

	We have examined the field dependence of the susceptibility by repeating the temperature scans shown in Fig.\  \ref{Chi-T} in a variety of applied fields, as is shown in Fig.\  \ref{Chi-T-B}.  Though we observed transient effects during the field sweeps, there was no significant hysteresis in the data.  Zero field cooling and field cooling produced the same values of $\chi'$ and $\chi''$.  Two aspects of the data set in Fig.\ 3 are readily apparent.  First, the small bump in $\chi'$ near 0.45 K has very little field dependence.  Second, is that the low temperature diamagnetic response is extinguished on a field scale that is small in comparison to $B_{c2}$ but large in comparison with $B_{c1}$.  Though the transport critical field in CePt$_3$Si is high, $B_{c2}\sim$ 4 - 5 T, and,  in fact, above the Clogston spin-paramagnetic limit  \cite{Clogston}, a field of only 0.5 T is sufficient to reduce the Meissner fraction by a factor of two.  On the other hand, the lower critical field of CePt$_3$Si is believed to be quite small.  The London penetration depth is estimated to be $\lambda_L\sim1$ $\mu$m and $\kappa\sim140$ \cite{CePt3Si}, which gives $B_{c1}\sim1$ mT  \cite{Tinkham}.  From this perspective it is surprising that a measurable Meissner fraction exists at a field $\sim500B_{c1}$ in a material with such a large value of $\kappa$. 
	
\begin{figure}
\includegraphics[width=6in]{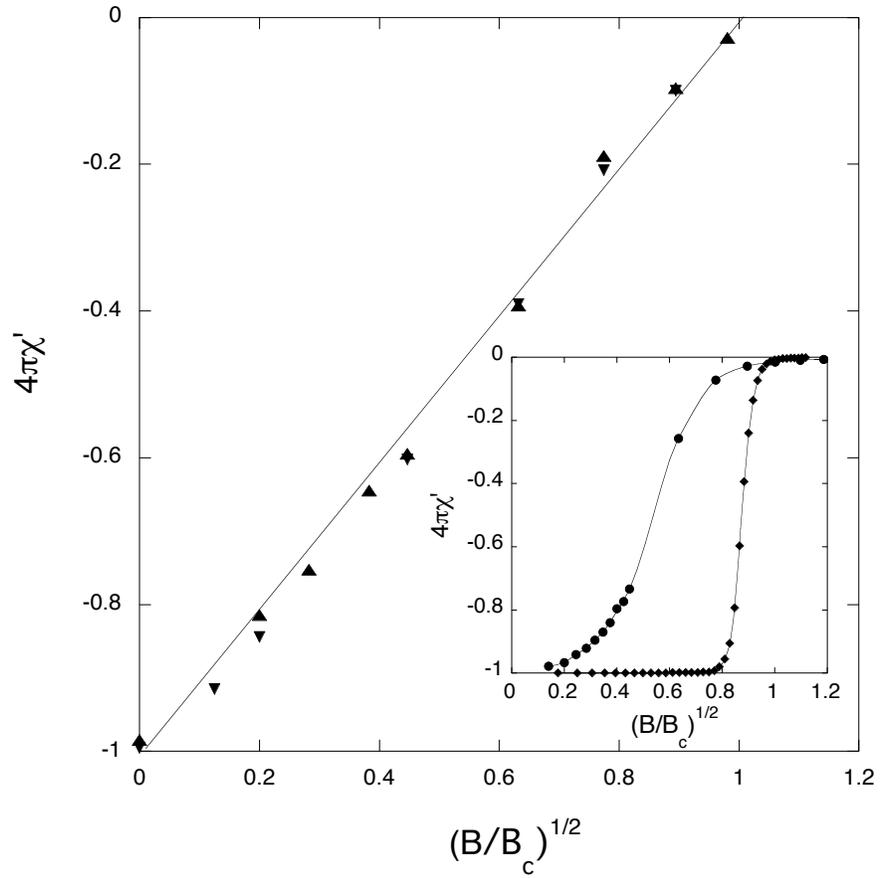}
\caption{\label{Chi-sqrtB} Square-root field dependence of the susceptibility of two CePt$_3$Si samples at 70 mK.  The line is provided as a guide to the eye.  Inset: the circles represent the 10 K susceptibility of the noncentrosymmetric $A15$ superconductor V$_3$Si.  The diamonds are data from a Pb sample alloyed with $5\%$ Sn in order to make it type II.  Note that neither of these latter systems obey a square-root scaling law. }
\newpage
\end{figure}

\begin{figure}
\includegraphics[width=6in]{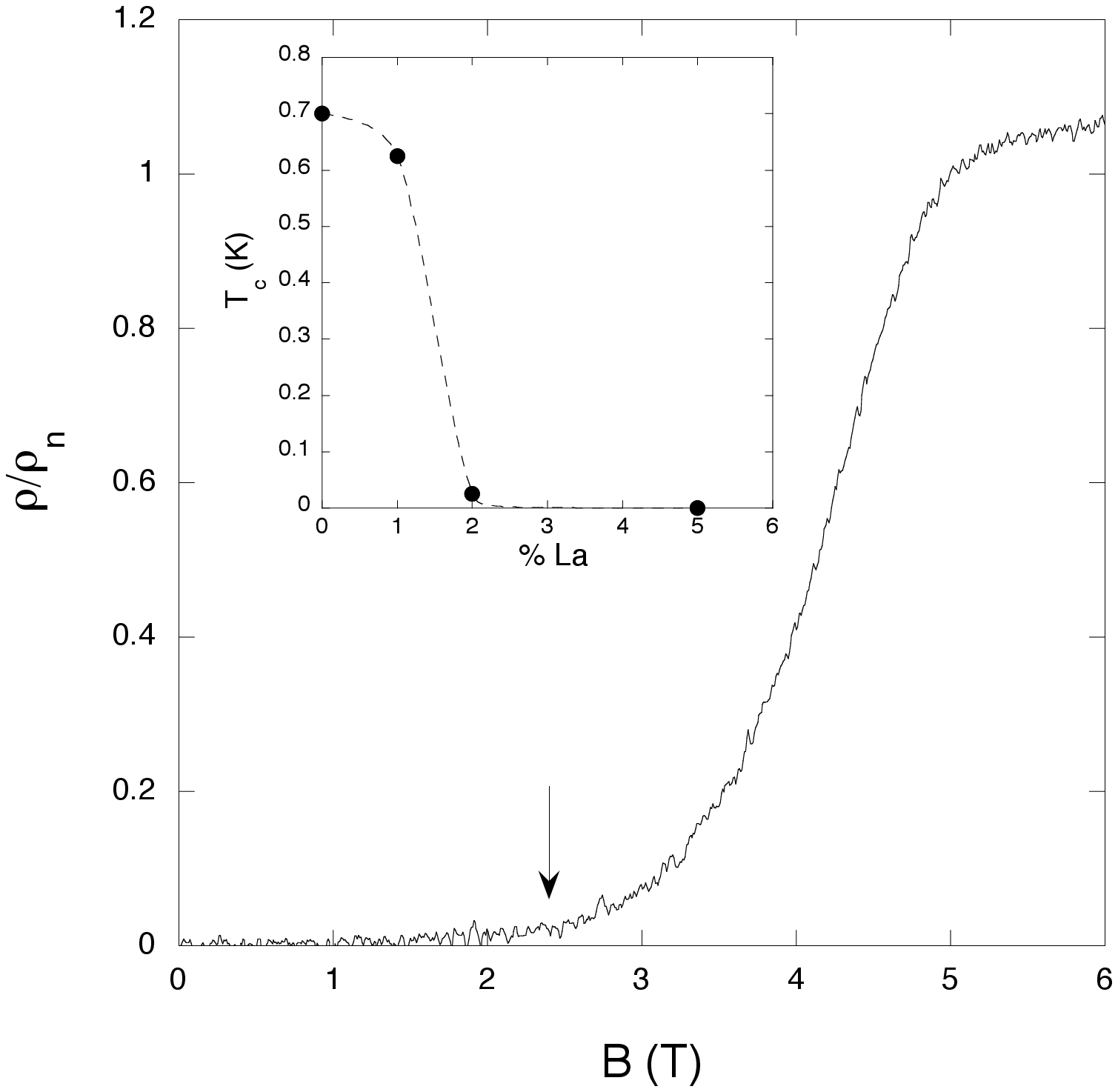}
\caption{\label{Rho-B}  Transport critical field transition in CePt$_3$Si.  The arrow depicts the field above which there is no longer a diamagnetic response in the susceptibility.  Note that this is also approximately the field at which the onset of resistance occurs.  Inset: transition temperature of Ce$_{1-x}$La$_{x}$Pt$_3$Si as a function of $x$.}
\newpage
\end{figure}

	The field dependence above $B_{c1}$ exhibits power-law behavior as can be seen in the main panel of Fig.\ \ref{Chi-sqrtB}, where the symbols represent the susceptibility of two different CePt$_3$Si samples.  In this plot we have scaled the field axis by the field at which the diamagnetic response goes to zero, $B_c=2.4$ T (see Fig.\ \ref{Rho-B}).  The linearity of the data demonstrates unambiguously that the low temperature susceptibility obeys a square root scaling law, $4\pi\chi'=(B/B_c)^{1/2}$ implying, of course, the same for the zero temperature susceptibility.  For comparison, we have also plotted the 10 K susceptibility of a polycrystalline V$_3$Si sample as a function of $B^{1/2}$ in the inset of Fig.\  \ref{Chi-sqrtB}.  V$_3$Si undergoes a second order structural phase transition just above its superconducting transition temperature $T_c=17$ K, which breaks the inversion symmetry of the crystal \cite{V3Si}.   The upper and lower critical fields are represented by the upper and lower corners of the curve, respectively \cite{Tinkham}.  Note the V$_3$Si does not obey the square-root scaling, indicating that the scaling is not simply a consequence of the absence of inversion symmetry.  For pedagogic purposes, we also have included in the inset of Fig.\ \ref{Chi-sqrtB} data from a Pb-5\%Sn sample which exhibits canonical type II behavior.  It is evident from a comparison between the data in the inset and the main body of Fig.\ \ref{Chi-sqrtB} that susceptibility of CePt$_3$Si is not representative of classic type II behavior and may be reflective of a nonconventional superconducting ground state \cite{Chi,Vekhter}.
	
	To the extent that itinerant antiferromagnetism mediates superconductivity in CePt$_3$Si, the critical field behavior may be driven by the field dependence of the underlying magnetic order parameter.  Additional evidence for the importance of the antiferromagnetic ground state in CePt$_3$Si can also be obtained via chemical substitution studies of La for Ce.  LaPt$_3$Si is isotypic to CePt$_3$Si but is not magnetic and does not superconduct.  By substituting non-magnetic La for Ce one should be able to disorder, and perhaps suppress, the N\'eel phase without significantly changing the crystal structure.  We have, in fact, studied several samples within the series Ce$_{1-x}$La$_x$Pt$_3$Si, where $x$ varied from 0 to 5\%.  Correct stoichiometry was confirmed via X-ray analysis.  Interestingly, quite small concentrations of La, $\sim2\%$, almost completely suppress the superconducting transition temperature, see the inset of Fig.\ \ref{Rho-B}.  The concurrent effect on the N\'eel transition is not completely known, but $\sim2\%$ La is enough to decrease $T_N$ below the base temperature of the PPMS magnetometer, 1.8 K.  The extraordinary sensitivity of $T_c$ to La doping, is comparable to what one would expect for magnetic impurity doping in a conventional superconductor.  A similar suppression of $T_c$ occurs in La doping of the 1-1-5's, but the transition temperature of Ce$_{1-x}$La$_x$CoIn$_5$, for instance, is an order of magnitude less sensitive to $x$ than what is shown in Fig.\ \ref{Rho-B} \cite{La115}.   We assume that the attenuation of $T_c$ is a pair breaking effect, though, of course, La is non-magnetic.  This would suggest that the pairing wavefunction is nonconventional in CePt$_3$Si, and that the enhancement of spin-orbit scattering in this material is producing a pronounced sensitivity to impurities.  
	
	In conclusion, the low temperature susceptibility of superconducting CePt$_3$Si exhibits a non-hysteretic, square-root power-law field dependence.  We believe that these two characteristics of the diamagnetic response are consistent with a nonconventional superconducting condensate and a nodal gap structure, in particular.   This conjecture is supported by the observed fragility of the superconducting phase to La substitution of Ce.  At this point,  however, the relative roles of antiferromagnetism, spin-orbit scattering, and the noncentrosymmetric crystal structure in pair formation and the subsequent expression of a macroscopic quantum state in CePt$_3$Si are unclear.    

	We gratefully acknowledge enlightening discussions with Dana Browne, Roy Goodrich, and Ilya Vekhter. This work was supported by the National Science Foundation under Grant DMR 02-04871.  JYC acknowldeges NSF-DMR-0237664 and Alfred P. Sloan Foundation for partial support of this project.


\newpage

\begin{references}

\bibitem{UGe2} S. Saxena {\it et al.}, Nature $\bf 406$, 587 (2000).

\bibitem{ZrZn2} C. Pfleiderer {\it et al.}, Nature $\bf 412$, 58 (2001).
	
\bibitem{CeCoIn5} H. Hegger {\it et al.}, Phys. Rev. Lett. $\bf 84$, 4986 (2000); C. Petrovic {\it et al.}, Europhys. Lett. $\bf 53$, 354 (2001); C. Petrovic {\it et al.} J. Phys. Condens. Matter $\bf 13$, L337 (2001).

\bibitem{CeRhIn5} S. Kawasaki {\it et al.}, Phys. Rev. Lett. $\bf 91$, 137001 (2003).

\bibitem{CePt3Si} E. Bauer {\it et al.}, Phys. Rev. Lett. $\bf 92$, 027003 (2004).

\bibitem{Inversion} L.P. Gorkov and E.I. Rashba, Phys. Rev. Lett. $\bf 87$, 037004 (2001);  S.K. Yip, Phys. Rev. B $\bf 65$, 144508 (2002).

\bibitem{Inversion2}  V. E. Edelstein, Phys. Rev. Lett. $\bf 75$, 2004 (1995); K.V. Samokhin, cond-mat/0405447 v1 (2004).

\bibitem{MnSi}  P.A. Frigeri, D. F. Agterberg, A. Koga, and M. Sigrist, Phys. Rev. Lett. $\bf 92$, 097001 (2004); V.P. Mineev, cond-mat/0405672 v1 (2004).

\bibitem{Neutron} N. Metoki {\it el al.}, J. Phys.: Condens. Matter $\bf 16$, L29 (2004).

\bibitem{QPT}  S. Sachdev, {\it Quantum Phase Transitions} (Cambridge Univ. Press, Cambridge, 1999).

\bibitem{Clogston} A.M. Clogston, Phys. Rev. Lett. $\bf 9$, 266 (1962); B.S. Chandrasekhar, Appl. Phys. Lett. $\bf
1$, 7 (1962).

\bibitem{Tinkham} M. Tinkam, {\it Introduction to Superconductivity} (McGraw-Hill, New York, 1996).

\bibitem{V3Si} P.W. Anderson and E.I. Blount, Phys. Rev. Lett. $\bf 14$, 217 (1965).

\bibitem{Chi} P.A. Frigeri, D.F. Agterberg, and M. Sigrist, cond-mat/0405179 v1 (2004); M. Yogi {\it et al.}, cond-mat/0405493 v1 (2004)

\bibitem{Vekhter} I. Vekhter, P.J. Hirschfeld, and E.J. Nicol, Phys. Rev. B $\bf 64$, 065512 2001.

\bibitem{La115} C. Petrovic, L. Bud'ko, V.G. Kogan, and P.C. Canfield, Phys. Rev. B $\bf 66$, 054534 (2002).

\newpage

\end{references}
\end{document}